\documentclass[aps,prb,twocolumn,floatfix,superscriptaddress]{revtex4-2}
\usepackage{graphicx,epsfig,array}
\usepackage{color,graphicx,amssymb,amsmath}
\usepackage[utf8]{inputenc}
\usepackage{physics}
\usepackage{tensor,upgreek}
\usepackage[dvipsnames]{xcolor}
\usepackage{color}
\usepackage{braket}
\usepackage{float}
\usepackage[colorlinks=true,linktoc=page,linkcolor=Blue,citecolor=Blue,urlcolor=Blue]{hyperref}

\newcommand\bea{\begin{eqnarray}}
\newcommand\eea{\end{eqnarray}}
\newcommand\beq{\begin{equation}}  
\newcommand\eeq{\end{equation}}

\newcommand{\non}{\nonumber}  

\newcommand\ie{{\it{i.e.}}}
\newcommand\etal{{\it{et al.}}}

\begin{document}
\textheight=23.8cm

\title{Correlation-Induced Topological Reconstruction in a Periodically Driven Kagome Mott Insulator}
\author{Rahul Ghosh}
 \email{rahul.ghosh@niser.ac.in}
\affiliation{School of Physical Sciences, National Institute of Science Education and Research, Jatni 752050, India}
\affiliation{Homi Bhabha National Institute, Training School Complex, Anushaktinagar, Mumbai 400094, India}

\author{Subhajyoti Pal}
\email{subhajyoti.pal@bose.res.in}
\affiliation{School of Physical Sciences, National Institute of Science Education and Research, Jatni 752050, India}
 \affiliation{Homi Bhabha National Institute, Training School Complex, Anushaktinagar, Mumbai 400094, India}
\affiliation{\mbox{S. N. Bose National Centre for Basic Sciences, Block JD, Sector III, Salt Lake, Kolkata 700106, India}}

\author{Ganesh C. Paul}
\affiliation{School of Physical Sciences, National Institute of Science Education and Research, Jatni 752050, India}
\affiliation{Homi Bhabha National Institute, Training School Complex, Anushaktinagar, Mumbai 400094, India}
\affiliation{Institut für Mathematische Physik, Technische Universität Braunschweig, D-38106 Braunschweig, Germany}


\begin{abstract}
We study the interplay between electronic correlations and circularly polarized periodic driving in the Hubbard model on a Kagome lattice. Using Brillouin--Wigner perturbation theory, we derive an effective Floquet Hamiltonian in which the drive renormalizes the bare hopping and generates chiral nearest- and next-nearest-neighbor terms, producing repeated topological transitions and a flat band whose position is continuously tunable across the spectrum. Within slave-rotor mean-field theory, we show that the resulting Mott transition is strongly non-monotonic in the drive amplitude, yielding repeated metal--insulator transitions, and that the spinon excitations inside the Mott phase acquire a band topology distinct from that of the non-interacting Floquet bands. This correlation-driven topological reconstruction produces emergent flat-band spinon insulators inaccessible to either driving or interactions alone. Our results establish periodically driven Kagome systems as a platform for engineering correlated topological flat-band physics out of equilibrium, with proposed realizations in ultracold atomic lattices.
\end{abstract}
\maketitle

\section{Introduction}
\label{sec:intro}

The interplay between electronic correlations and nontrivial band topology has become a central paradigm in condensed matter physics, driving the discovery of exotic equilibrium phases such as topological insulators~\cite{ti-rmp}, topological Mott insulators~\cite{Pesin_Balents,PhysRevLett.100.156401}, axionic phases~\cite{axion}, and quantum spin liquids~\cite{Balents2010,Savary2017,Zhou2017}. Parallel to these equilibrium advances, periodic electromagnetic driving has emerged as a powerful out-of-equilibrium route for engineering quantum matter, establishing Floquet physics as a major research frontier~\cite{oka2009photovoltaic,lindner2011floquet,dora2012optically}, spearheaded by landmark experiments in photonic systems~\cite{rechtsman2013photonic}, quantum materials~\cite{wang2013observation}, and ultracold atomic lattices~\cite{jotzu2014experimental,peng2016experimental}. Floquet engineering enables the controlled modification of electronic band structures~\cite{sambe1973steady,klinovaja2016topological}, the generation of synthetic gauge fields, the realization of non-equilibrium Majorana modes~\cite{jiang2011majorana,kundu2013transport,perez2014floquet}, novel transport phenomena~\cite{gu2011floquet,kundu2014effective,farrell2015photon}, and disorder-driven effects~\cite{titum2015disorder} inaccessible in equilibrium.

A fundamental question at the intersection of these two directions is whether electronic correlations simply renormalize the topology of Floquet-engineered bands or qualitatively reconstruct it. While periodic driving bestows non-trivial band topology on otherwise trivial systems through virtual photon processes in the non-interacting limit~\cite{casas2001floquet,blanes2009magnus,mikami2016brillouin}, much less is understood about how strong correlations modify these non-equilibrium topological states. This question sits at the boundary of two independently active frontiers—-Floquet engineering and strongly correlated Mott physics—-and its resolution requires a material platform where the drive simultaneously generates multiple independent chiral hopping channels that correlations can treat differently.

Two scales govern the response of an interacting system to a light drive: the driving frequency $\hbar\Omega$ and the on-site interaction $U$. Their relative magnitude defines three distinct regimes: low-frequency ($\hbar\Omega \ll U$), intermediate ($\hbar\Omega \sim U$), and high-frequency off-resonant ($\hbar\Omega \gg U$). Strong-coupling spin models~\cite{PhysRevB.99.205111}, Kondo insulators~\cite{PhysRevB.96.115120}, and multi-orbital Hubbard models~\cite{PhysRevB.99.205111} have all been studied in the presence of a light drive, and the off-resonant driving of Bose-Hubbard models has been explored both experimentally~\cite{PhysRevLett.115.155301} and theoretically~\cite{PhysRevB.92.125107, PhysRevLett.109.203005,PhysRevLett.103.133002}. Among fermionic correlated systems, the  honeycomb, square and triangular lattices have served as prototypical testbeds~\cite{jana2020tailoring,flint-prb}. Strikingly, in both cases the topology of the emergent spinon excitations in the Mott phase \textit{faithfully inherits} that of the non-interacting Floquet bands. Whether this correspondence is a universal property of driven correlated fermions, or whether it breaks down in geometrically frustrated systems with richer hopping structures, remains an open and unresolved question.

The Kagome lattice provides a particularly attractive platform for addressing this problem, for three reasons. First, its geometry natively hosts a perfectly flat band arising from destructive interference on triangular plaquettes~\cite{Bergman2008,TIkago,PhysRevLett.106.236804}, making the electronic structure exceptionally sensitive to perturbations of the hopping network.
Second, under off-resonant circularly polarized light, the high-frequency BW expansion generates not one but \textit{two} independent emergent hopping channels: chiral NN hopping $t^0_{ij}$ and chiral NNN hopping $t^1_{ij}$~\cite{mikami2016brillouin,PhysRevB.82.235114}. This two-channel structure is absent in the triangular lattice~\cite{jana2020tailoring}, where only the NNN channel appears, and provides the microscopic origin of the breakdown of topology inheritance. Third, the Kagome flat band enhances correlation effects by suppressing kinetic energy, making the Mott transition and the associated topological reconstruction accessible at moderate values of $U/t$.

In this work, we study the Hubbard model on the Kagome lattice subjected to off-resonant circularly polarized light at half filling. Using BW perturbation theory, we derive the effective Floquet Hamiltonian comprising, renormalized NN, emergent chiral NN, and emergent chiral NNN hopping channels, and show that these produce repeated topological transitions and a spectral flat band whose position is continuously tunable across the three-band structure as a function of drive amplitude. To incorporate strong electronic correlations, we employ slave-rotor cluster mean-field theory (SR-CMFT)~\cite{florens2004slave,zhao2007self} in the paramagnetic sector. We demonstrate that the critical interaction strength $U_C$ for the Mott transition is strongly non-monotonic in the drive amplitude $A_0$, yielding repeated metal--insulator transitions driven by periodic suppression of the kinetic energy. Most importantly, we show that inside the Mott phase, the NN and NNN hopping channels are renormalized by the rotor correlators \textit{unequally}: because the NNN channel is entirely drive-generated while the NN channel retains a bare contribution, interactions preferentially suppress the former and independently modify the complex phases of both. This channel-selective phase renormalization provides the microscopic mechanism for reconstructing the effective gauge flux through elementary hopping loops, producing spinon Chern numbers distinct from those of the non-interacting Floquet bands---a qualitative topological reconstruction that is absent in previously studies~\cite{jana2020tailoring}. These results suggest that topology inheritance need not be a universal property of Floquet Mott systems but depends sensitively on the structure of the underlying Floquet Hamiltonian. One notable consequence of this reconstruction is the emergence of \textit{topological spinon flat bands} deep inside the Mott insulating phase, coexisting with a robust charge gap and tunable through the drive amplitude. We distill this physics into a minimal effective chiral Hamiltonian that provides a minimal theoretical framework for engineering correlated topological flat-band states.

A practical concern for any Floquet proposal is heating. Periodic driving of isolated interacting systems generically leads to a featureless
infinite-temperature state at long times~\cite{PhysRevX.4.041048}. However, in the off-resonant regime, the heating rate is exponentially suppressed as
$\sim e^{-\Omega/t}$~\cite{abanin2017rigorous}, giving rise to a long-lived prethermal window~\cite{PhysRevLett.115.256803,Abanin2017,PhysRevE.90.012110,Weidinger2017} in which
the effective Floquet Hamiltonian provides an accurate description of the dynamics on experimentally relevant timescales. The high-frequency regime considered here is expected to lie within this long-lived prethermal window.

The paper is organized as follows. Section~\ref{sec_II} introduces the model and derives the effective Floquet Hamiltonian within BW perturbation theory, followed by the slave-rotor mean-field framework. The non-interacting Floquet band structure and the interacting phase diagram---including the correlation-induced topological reconstruction and emergent spinon flat bands---are presented in Sec.~\ref{sec_III}. Experimental feasibility is discussed in Sec.~\ref{sec_IV}, and we conclude with a broader perspective in Sec.~\ref{sec_V}.

\section{Model and Method} \label{sec_II}

\begin{figure}
	\includegraphics[width=0.7\linewidth]{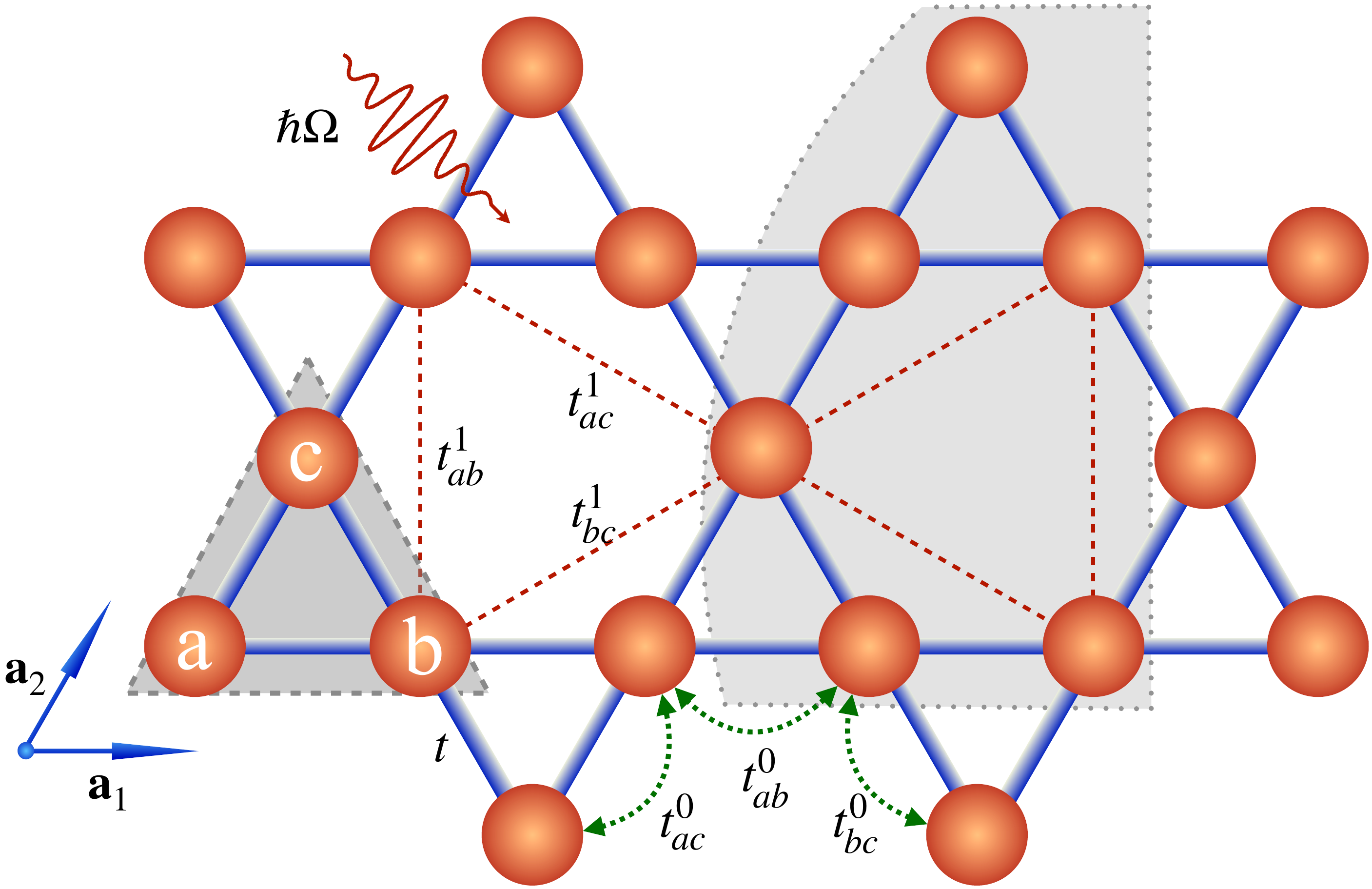}
	\caption{Schematic diagram of the kagome lattice, consisting of three atoms, $a$, $b$, and $c$, in each unit cell (shaded triangle). Here, $t_{i,j}$ $(i,j\in\{a,b,c\})$ denotes the nearest-neighbour (NN) hopping amplitude between the three atoms (solid lines), while $t_{ij}^0$ $(i,j\in\{a,b,c\})$ denotes the emergent nearest-neighbour complex hopping induced by the external drive (green dashed lines). Emergent next-nearest-neighbour (NNN) complex hopping is represented by $t_{ij}^1$ $(i,j\in\{a,b,c\})$ (red dashed lines). The small shaded triangle represent the unit cell structure and the large shaded region represents the six-site cluster used in the slave-rotor mean-field theory.}
	\label{fig-1}
\end{figure}

We consider Kagome lattice structure where $a$, $b$, $c$ are the sub-lattices as shown in the Fig. $\ref{fig-1}$. 
In the absence of interaction, the tight binding Hamiltonian $H_{Free}$ can be written as 
\bea
&&H_{Free}=-\sum_{i,j,\sigma }(t_{ij}^{} d_{i\sigma }^{\dagger}d_{j\sigma }+h.c)
\label{e-1}
\eea

Here, $t_{ij}^{}\in \{t_{ab},t_{bc},t_{ca}\}$ are hopping amplitudes connecting nearest neighbour site pairs $a-b, b-c, c-a$ respectively. $d^\dagger_{i\sigma}$ $(d_{i,\sigma})$ is the creation (annihilation) operator of a electron with spin $\sigma$ at a site $i$. 
\subsection{Light drive:} 

For deriving the effective Hamiltonian in presence of a light drive, we choose three-site primitive (triangular) unit-cell containing one site from each sub-lattice as demarcated by the shaded triangle in Fig. $\ref{fig-1}$. The primitive unit vectors,  $\mathbf{a}_1=(a,0),\,\mathbf{a}_2=(a/2,\sqrt{3}a/2),\,\mathbf{a}_3=\mathbf{a}_2-\mathbf{a}_1=(-a/2,\sqrt{3}a/2)$ where $a$ is the inter-atomic distance, are shown in the Fig.~\ref{fig-1}. 
We couple to this lattice a circularly polarised light of the form $\mathbf{A}=A_0(\cos{(\Omega \tau)},\sin{(\Omega \tau)})$ where amplitude $A_0$ and frequency $\Omega$ via usual Peierls substitution $t^{}_{ij}\rightarrow t^{}_{ij} e^{-iA_0a\cos{(\Omega \tau-\pi l/3)}}$  where $l = 0,1,2$ for the three site-pairs $a-b, b-c, c-a$ respectively. In what follows we set $|t_{ij}|\equiv t$ , for all $i,j$. 

Several methods exist \cite{casas2001floquet,blanes2009magnus,eckardt2015high} for deriving the effective Hamiltonian of a periodically driven system in the high-frequency limit, $\ie$, when the driving frequency is much larger than the bandwidth ($\Omega\gg t$). We use Brillouin--Wigner (BW) perturbation theory~\cite{mikami2016brillouin} in the zero-photon sector (see appendix~\ref{BW}). We choose circularly polarised light instead of linearly polarised light because it explicitly breaks time-reversal symmetry through its finite helicity. Each photon carries angular momentum $\pm\hbar$ along the propagation direction. Under time reversal, the photon helicity changes sign, converting a right-handed circularly polarised field into a left-handed one. Since the external drive has a fixed helicity, the driven Hamiltonian is not time-reversal invariant, $\mathcal{T}H(\tau)\mathcal{T}^{-1}\neq H(\tau)$. As a result, the virtual photon absorption and emission processes captured by the BW expansion generate direction-dependent complex hopping amplitudes, which act as an effective gauge field or synthetic magnetic flux in the electronic Hamiltonian. These emergent chiral hopping terms are responsible for the non-trivial Floquet band topology obtained in the high-frequency limit~\cite{arijitSilicene,usaj2014irradiated,perez2014floquet}.

A time periodic Hamiltonian $H(\tau+T)=H(\tau)$ given by its Fourier components as
\begin{align}
\mathcal{H}_{p} = \int_{0}^{T}\frac{d\tau}{T} H(\tau) e^{ip \Omega \tau}\ , \label{e-2}
\end{align}
where $T=2\pi/\Omega$ is the time period of drive. We closely following Mikami \etal~\cite{mikami2016brillouin} and obtain the effective Hamiltonian in powers of $1/\Omega$ using the BW perturbation theory. The effective Hamiltonian can be written as
\begin{equation}
\mathcal{H}_{\textrm{BW}}=\mathcal{H}_0+ \sum_{p \neq 0} \frac{\mathcal{H}_{-p}\mathcal{H}_p}{p\Omega} \ ,
\label{e-3}
\end{equation}
up to order $1/\Omega$, where we neglect terms of order $1/\Omega^2$ and higher. The summation over $p$ denotes virtual $p$-photon excitations, as defined in the seminal work of Sambe \cite{sambe1973steady}.
We extract the essential physics by truncating the Hamiltonian at order $1/\Omega$ in the high-frequency limit. The zeroth-order Hamiltonian $\mathcal{H}_{0}$ describes the unperturbed system with renormalised hopping parameters, i.e., $H_0=J_0(A_0) H_{Free}$, where $J_0(A_0)$ is the Bessel function of the first kind of order zero. The $O(1/\Omega)$ terms arise from the periodic drive and generate hopping processes~\cite{mikami2016brillouin, arijitSilicene,usaj2014irradiated}; these are given by the second and third terms in Eq.\ref{e-4}, which defines the effective Hamiltonian.
\bea
\mathcal{H}_{BW}&=J_0(A_0)H_{Free}+\sum_{\langle i,j\rangle}t\chi_1 \nu_{i,j}(d^{\dagger}_{i}d_j+h.c.)\nonumber\\
    &+\sum_{\langle\langle i,j\rangle\rangle}t\chi_2 \nu_{i,j}(d^{\dagger}_{i}d_j+h.c.)
	\label{e-4}
\eea
$\nu_{ij}=\pm 1$ for clockwise (+1) and anticlockwise (-1) hopping for both spin channels and $\chi_1=\sum_p \frac{2i}{p\omega}(-1)^psin(p\pi/3)J_p(A_0)$ and $\chi_2=\sum_p \frac{2i}{p\omega}sin(p\pi/3)J_p(A_0)$. The details of the derivation of $\mathcal{H}_{BW}$ is provided in the Appendix-~\ref{BW}. The chirality in the emergent NN and NNN terms leads to spin-orbit coupling like term in the spin-orbit coupling in the Haldane model. In the above, we have suppressed the spin index of the operators as the analysis is independent of the spins. We now reinstate the spin indices to analyse interaction effects. As discussed in the Appendix~\ref{BW}, we choose a triangular unit cell, comprising of one atom from each sub-lattice for calculating the bands. We note that in general driving an isolated many-body system  causes heating effects and eventually leads to a featureless thermal state~\cite{PhysRevX.4.041048}. However, it is well-established that there is a `pre-thermal' regime where the light-driven Hamiltonian is well-defined \cite{prethermal}. However, it has been shown that the time scale of heating is quasi exponentially slow~\cite{PhysRevLett.115.256803,Abanin2017} allowing one to work in a pre-thermal regime which survives for experimentally relevant time scales~\cite{Weidinger2017}.

\begin{figure*}[t]
	\hspace*{\fill}%
	\includegraphics[width=1\linewidth]{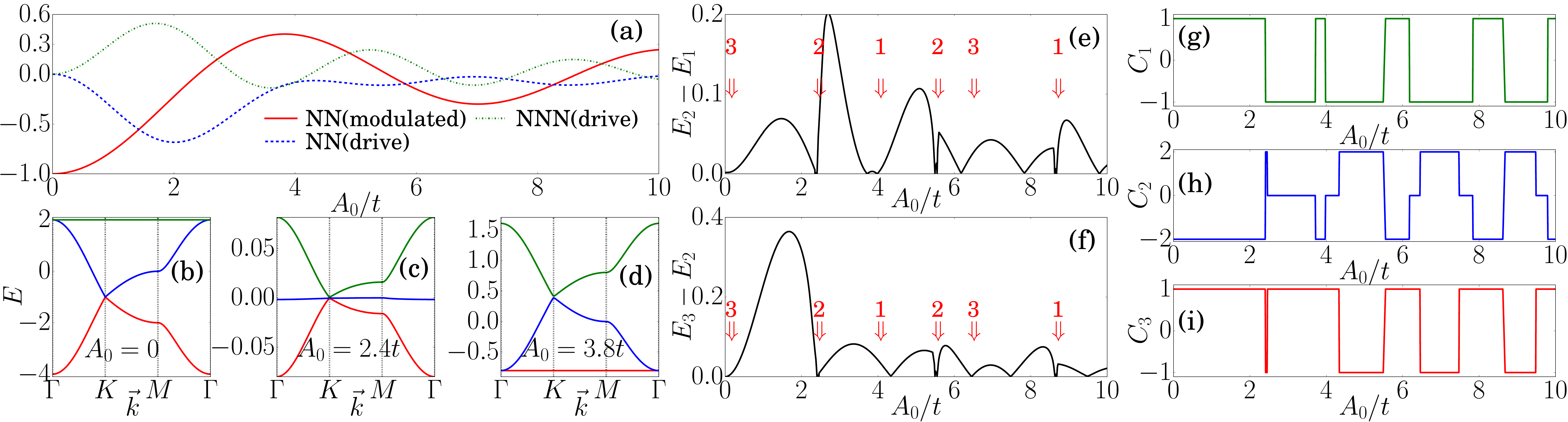}\hfill
	\hspace*{\fill}%
	\caption{(Color online) Upper panel (a) shows the dependence of the modulated real part of NN hopping ($t^R_{ij}$), the drive-generated imaginary part of  NN hopping ($t^0_{ij}$) and emergent NNN hopping ($t^1_{ij}$) as a function of the driving amplitude $A_0$. The drive-generated NN and NNN hopping amplitudes are small, So we plotted these  two hopping multiplying with 20. Lower panel [(b) to (d)] shows the non-interacting tight binding bands of driven kagome lattice for $A_0 =0,\,2.4t,\,3.8t$ respectively. For $A_0=0$, the top band (green) is flat, while for $A_0/t=2.4$ the middle band (blue) gets flattened, whereas the bottom band (red) becomes flat for $A_0/t=3.8$. In the middle (e) and (f) panels evolution of the band gap is shown as a function of driving amplitude $A_0/t$. The evolution of the band gap between the bottom and the middle band is shown in panel (e), and the band gap between the top and middle band is shown in (f) panel. The right three panel [(g)-(i)] shows the change of the Chern number as a function of driving amplitude $A_0/t$. Chern number changes in every band touching point.}
	\label{fig-2}
\end{figure*}
\subsection{Interaction effects} 
We include local interaction between the electrons in $\mathcal{H}_{BW}$. The resulting $H_{full}$ defined as: 
\begin{eqnarray}
&& H_{Full}=\mathcal{H}_{BW}  +U\sum_i n_{i\uparrow}n_{i\downarrow}
	\label{e-5}
\end{eqnarray}  
We note that in the off-resonant regime the effective interacting Hamiltonian indeed retains the above mentioned form as discussed in literature \cite{ham-form-1,ham-form-2}. We employ slave-rotor mean field theory to study the interaction effect in the off-resonant limit. For this we closely follow the slave-rotor decomposition as described in \cite{florens2004slave,zhao2007self,Ghosh2025} and rewrite the creation and annihilation operators as:
\begin{equation}
d_{i\sigma}^{\dagger} = f_{i\sigma}^{\dagger}\,e^{-i\theta_i},
\qquad
d_{i\sigma} = f_{i\sigma}\,e^{+i\theta_i},
\label{e-6}
\end{equation}
where $f^\dagger_{i\sigma}$ denotes the auxiliary fermion (spinon) operator at the $i^{\rm th}$ site and $e^{ i\theta_{i}}$ ($e^{ -i\theta_{i}}$) represent the rotor creation (annihilation) operator which is defined as: $e^{\pm i\theta_{i}}|n_{i}^{\theta}\rangle=|n_{_{i}}^{\theta}\pm1\rangle$. The transformation imposes a direct product structure at each site of the lattice, (i) a spinon that keeps track of the fermionic anti-commutation and (ii) a rotor that tracks the charge degrees of freedom. However, caution must be exercised in going forward as this direct product space contains unphysical states. As is well known the standard way out is to impose a suitable local constraint $n_{i}^{\theta}+n_{i\uparrow}^{f}+n_{i\downarrow}^{f}=1$ at half filling. Here, $n_{i\sigma}^{f}=f^\dagger_{i\sigma}f_{i\sigma}$, with $n_{i\sigma}^{f}=n_{i\sigma}^{e}$. As it is well-known that this constraint exactly projects out the unphysical states. The discussion so far has been for an exact slave-rotor analysis. 

Here we follow a mean-field ansatz that the ground state of $H_{Full}$, $|\Psi\rangle$ is a direct product to the spinon and rotor ground states, \textit{i.e.} $|\Psi\rangle=|\Psi^{f}\rangle\otimes|\Psi^\theta\rangle$. The solution then proceeds by self-consistently solving two coupled Hamiltonians, $H_{\theta}\equiv \langle \Psi^{f}|H_{Full}|\Psi^{f}\rangle$ and $H_{f}\equiv \langle \Psi^{\theta}|H_{Full}|\Psi^{\theta}\rangle$. As detailed in the Appendix~\ref{SL-formalism}, $H_{f}$ has the same form as $\mathcal{H}_{BW}$ and is a non-interacting Hamiltonian. $H_{\theta}$ on the other hand, is a problem of itinerant rotor degrees of freedom along with local rotor repulsion. We solve the later using a cluster mean-field theory, for which we choose a six-site cluster as depicted in the Fig.~\ref{fig-1}. An exact diagonalisation of the cluster allows calculation of renormalisation factors for the spinon-Hamiltonian. Since the spinon-Hamiltonian contains NN and NNN hopping a minimum of six site cluster (shaded part of Fig.~\ref{fig-1}) needs to be chosen to calculate all six hopping renormalisation factors . The above-mentioned constraint is imposed on the cluster as $\sum_{i=1}^6(n_{i}^{\theta}+n_{i\uparrow}^{f}+n_{i\downarrow}^{f}-1)=0$. The constraint is satisfied at a mean-field level within our formalism as discussed in the Appendix~\ref{SL-formalism}. We emphasise that the symmetry of the NN and NNN hopping amplitude renormalisation factors remarkably allows us to still use a three-site unit cell for solving $H_f$ as for the non-interacting problem.

\textit{\underline{Observables}:} Within the slave-rotor-mean-field-theory (SR-MFT), we calculate $\langle\Psi^{\theta}| e^{-i\theta_{i}}|\Psi^{\theta}\rangle \equiv\Phi_{i}$ which is mean field order parameter with $i$ running over all six sites within the rotor-cluster. $\Phi_{i}=0$  $\forall i$ \cite{florens2004slave,zhao2007self} indicates a Mott insulating phase, on the other hand, $\Phi_{i}\neq 0$ indicates a metallic phase. To investigate metal-insulator transition, we calculate the sub-lattice local projected density of states (PDOS) defined as $N_{i}(\omega)$ that provides the spectral weight at  site $i$. It s expression in terms of the rotor and spinon one-body Green's-functions is provided in the Appendix~\ref{DOSAPP}. We also calculate band Chern numbers for the non-interacting bands, and for the spinon-band in the interacting case from the solution of $H_f$.
\begin{figure*}[t]
	\includegraphics[width=1\linewidth]{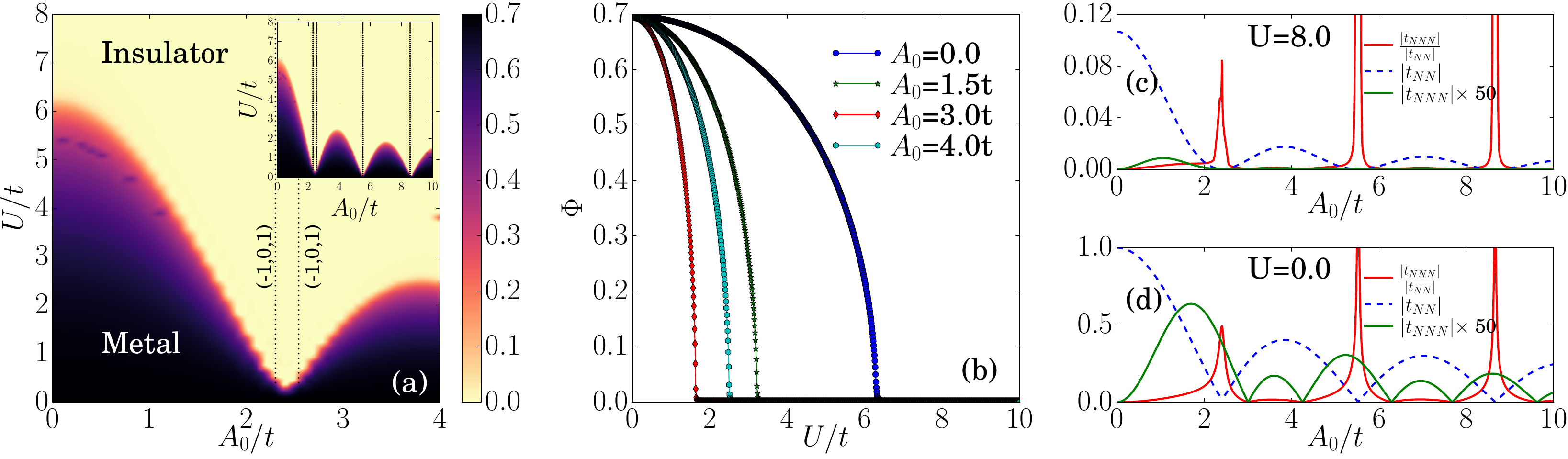}

\caption{(a) Ground-state phase diagram in the $U$--$A_0$ plane obtained within SR-CMFT. The dark and yellow regions correspond to the metallic and Mott insulating phases, respectively, separated by the critical interaction strength $U_C$. The dotted vertical lines indicate the driving amplitudes at which the middle spinon band becomes flat; the corresponding Chern numbers are shown. The inset illustrates the repeated metal--insulator transitions over a wider range of driving amplitudes. (b) Evolution of the rotor order parameter, $\Phi = \tfrac{1}{6}\sum_{i=1}^{6}\Phi_i$, as a function of $U$ for the driving amplitude $A_0=$0, 1.5t, 3.0t, and 4.0t, demonstrating the non-monotonic dependence of $U_C$ on the drive strength. (c),(d) Evolution of the magnitudes of the effective NN hopping, $|t|$, the NNN hopping, $|t_{ij}^{1}|$, and their ratio $|t_{ij}^{1}|/|t|$ as functions of the driving amplitude correspond to the interacting ($U=8$) and non-interacting ($U=0$) cases, respectively.}

	\label{fig-3}
\end{figure*}
\section{Results}\label{sec_III}

\subsection{Non-interacting driven system}

We begin by investigating the effect of the periodic drive in the high-frequency off-resonant regime. Throughout this work, we set the nearest-neighbour (NN) hopping amplitude $t=1$, the driving frequency $\Omega=20$, and perform all calculations at zero temperature and half filling.

The influence of the periodic drive on the effective hopping amplitudes is summarized in Fig.~\ref{fig-2}(a). Within the BW formalism, the bare NN hopping is renormalized according to $t^{R}_{ij}=J_{0}(A_{0})t$ $\forall i,j$, where $J_{0}$ is the zeroth-order Bessel function. Consequently, the effective hopping vanishes at the zeros of $J_{0}(A_{0})$. In addition to this bandwidth renormalization, virtual photon absorption and emission processes generate complex NN ($t_{ij}^{0}$) and NNN ($t_{ij}^{1}$) hopping amplitudes. Although these drive-induced hopping terms are smaller in magnitude than the renormalized bare hopping, they break time-reversal symmetry and are responsible for the emergence of non-trivial Floquet band topology.

The effective Hamiltonian is diagonalized in momentum space using a three-site unit cell, yielding three Floquet bands. The corresponding band structures along the high-symmetry path $\Gamma-K-M-\Gamma$ are shown in Fig.~\ref{fig-2}(b)--(d). In the absence of the drive ($A_{0}=0$), the Kagome lattice exhibits the familiar electronic structure consisting of a completely flat upper band and two dispersive bands touching at the $K$ point [Fig.~\ref{fig-2}(b)]. As the driving amplitude increases, the competition between the renormalized bare hopping and the drive-induced chiral hopping processes leads to a substantial reconstruction of the band structure.

One of the most remarkable consequences of the periodic drive is the ability to continuously tune the position of the flat band within the spectrum. At $A_{0}/t=2.4$, the middle band becomes nearly dispersionless [Fig.~\ref{fig-2}(c)], while at $A_{0}/t=3.8$ the flat band is transferred to the bottom of the spectrum [Fig.~\ref{fig-2}(d)]. The evolution of the flat-band position with the driving amplitude is highlighted by the red arrows in Fig.~\ref{fig-2}(e) and Fig.~\ref{fig-2}(f), demonstrating that the flat band oscillates periodically between the upper, middle, and lower bands. This provides a simple and controllable mechanism for engineering topological flat bands in the Kagome lattice and is expected to play a significant role in the interacting regime discussed below.

The evolution of the band topology is characterized by the energy gaps between adjacent bands, $E_{2}-E_{1}$ and $E_{3}-E_{2}$, shown in Fig.~\ref{fig-2}(e) and Fig.~\ref{fig-2}(f), respectively. Both gaps undergo repeated closing and reopening as the driving amplitude is increased, signaling a sequence of topological phase transitions. The locations of these gap closings closely coincide with the zeros of the any one of the three light induced hopping amplitude, indicating that the periodic suppression of the kinetic energy governs the band inversions.

To establish the topological nature of these transitions, we calculate the Chern numbers of all three Floquet bands, shown in Fig.~\ref{fig-2}(g)--(i). We find that the Chern numbers change precisely at the band-touching points where either $E_{2}-E_{1}$ or $E_{3}-E_{2}$ vanishes, confirming that every gap closing corresponds to a topological phase transition between distinct Floquet Chern insulating phases. Therefore, periodic driving not only provides a powerful route for engineering the bandwidth and the position of the flat band, but also enables repeated topological band inversions, giving rise to a rich sequence of topological phases that can be tuned continuously by the driving amplitude.
\subsection{Interaction effects}

We now investigate the effect of electronic correlations on the periodically driven Kagome lattice within slave-rotor cluster mean-field theory (SR-CMFT). Figure~\ref{fig-3}(a) shows the resulting ground-state phase diagram in the $U$--$A_0$ plane, with the color scale representing the magnitude of the rotor order parameter $\Phi = \tfrac{1}{6}\sum_{i=1}^{6}\Phi_i$, averaged over the six-site rotor cluster [shaded region, Fig.~\ref{fig-1}]. All six rotor order parameters vanish simultaneously at the Mott transition, ruling out any site-selective Mott physics. Since $\Phi$ measures the coherence of the charge degrees of freedom, $\Phi \neq 0$ identifies a metallic state while $\Phi = 0$ signals the freezing of charge fluctuations and the onset of the Mott insulating phase. Figure~\ref{fig-3}(b) shows the evolution of $\Phi$ with $U$ for several driving amplitudes; the critical interaction strength $U_C$ depends strongly and non-monotonically on $A_0$, demonstrating that the periodic drive offers an efficient handle on the correlation-induced metal--insulator transition.

Although the local rotor condensate vanishes in the Mott phase, the inter-site rotor correlations $Q_{ij}=\langle e^{-i\theta_i}e^{i\theta_j}\rangle$ remain finite and decay approximately as $1/U$. These correlations capture virtual charge fluctuations that survive inside the Mott state---one of the principal advantages of the slave-rotor approach over conventional static mean-field theory---and they renormalize the hopping amplitudes entering the spinon Hamiltonian, thereby controlling the low-energy spin physics. Figures~\ref{fig-3}(c) and \ref{fig-3}(d) show the renormalized NN and NNN hopping amplitudes as functions of $A_0$ for $U=8$ and $U=0$, respectively. The ratio $|t_{ij}^{1}|/|t|$ diverges whenever the NN hopping approaches zero and vanishes whenever the NNN hopping vanishes; notably, the two never vanish simultaneously. The NNN hopping is far more strongly suppressed by correlations than the NN hopping, since it is generated entirely by the drive, whereas the NN channel retains a bare contribution alongside its drive-induced piece.

This strong renormalization of the hopping amplitudes is directly reflected in the phase diagram: $U_C$ drops [Fig.~\ref{fig-3}(b)] sharply wherever the effective NN hopping is small, since correlation effects are amplified as the kinetic energy is suppressed by the drive. The system therefore exhibits repeated metal--insulator transitions as a function of $A_0$ [inset, Fig.~\ref{fig-3}(a)], with the overall decay of the $U_C$ maxima tracking the drive-induced reduction of the effective bandwidth.

\begin{figure}[t]
	\hspace*{\fill}%
	\includegraphics[width=0.8\linewidth]{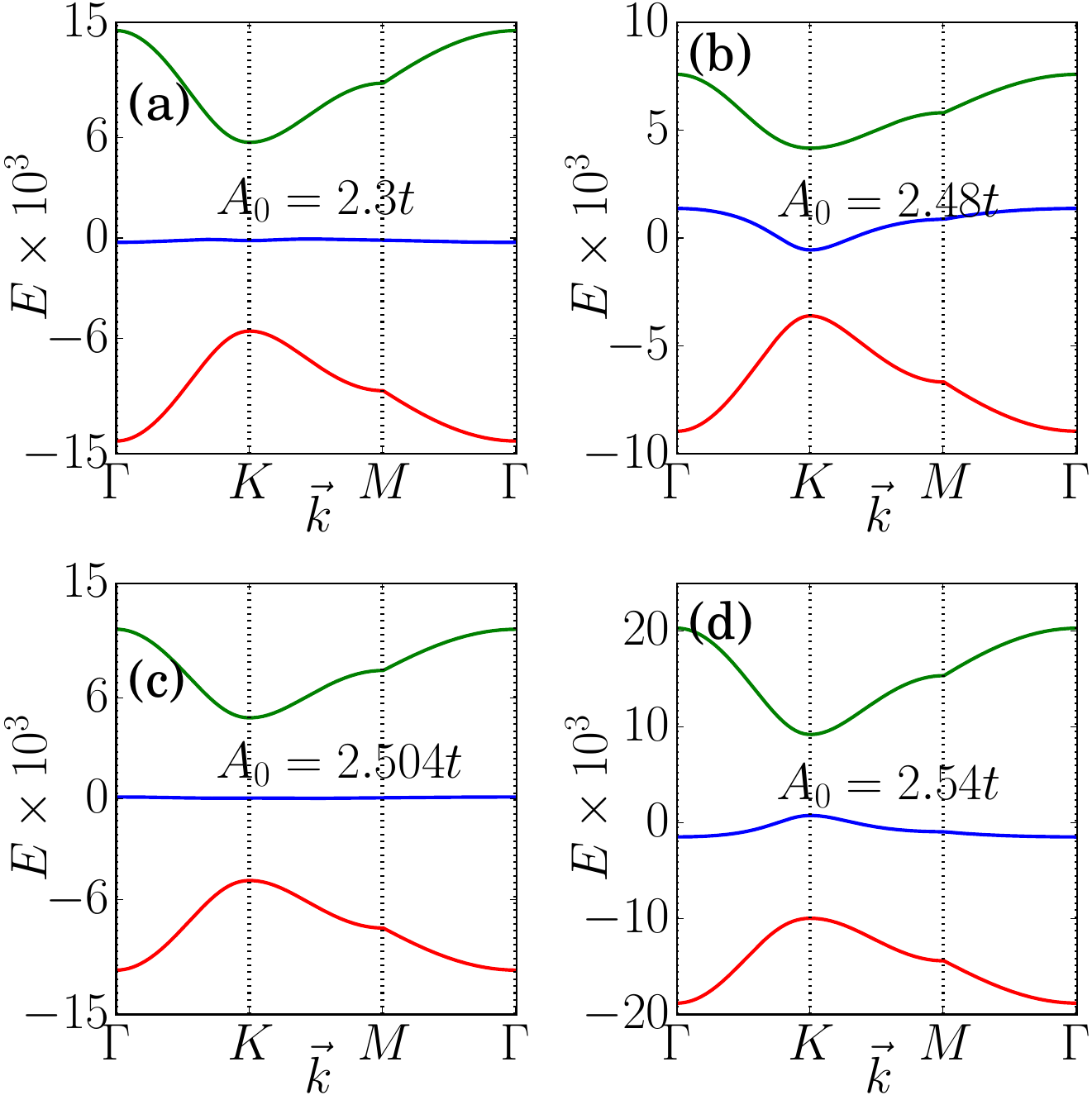}
	\hspace*{\fill}%
     \caption{(Color online) Spinon bands for different values of the driving amplitude $A_0$ at $U=8t$. The middle band is flat in panels (a) and (c), corresponding to driving amplitudes of $A_0 = 2.3t$ and $A_0 = 2.504t$, respectively. Conversely, panels (b) and (d) display dispersive bands for $A_0 = 2.48t$ and $A_0 = 2.54t$, respectively.}
	\label{fig-4}
\end{figure}

A central consequence of the slave-rotor formalism is that, although the spinon Hamiltonian retains the same operator structure as the non-interacting Floquet Hamiltonian, each hopping amplitude is renormalized by the corresponding rotor correlation function,
$Q_{ij}=\langle e^{-i\theta_i}e^{i\theta_j}\rangle$. Since the effective Floquet hopping amplitudes are complex, the renormalized spinon hoppings generally acquire both modified magnitudes and phases. Consequently, the effective gauge field experienced by the spinons is also renormalized. For an elementary triangular plaquette, the accumulated phase is
\begin{equation}
\phi_{\triangle}
=
\arg(Q_{ab}t_{ab}^{\rm eff})
+
\arg(Q_{bc}t_{bc}^{\rm eff})
+
\arg(Q_{ca}t_{ca}^{\rm eff}),
\label{eq:flux}
\end{equation}
which provides an illustrative measure of the effective gauge flux associated with the triangular hopping loop. Since the NN and NNN hopping channels acquire different renormalization factors, the effective flux experienced by the spinons generally differs from that of the non-interacting Floquet Hamiltonian. This modifies the Berry-curvature distribution of the spinon bands and consequently leads to a different sequence of Chern numbers.

The correlation-induced topological reconstruction uncovered here has no counterpart in the periodically driven triangular lattice, where the spinon and electronic Floquet band topologies remain identical~\cite{jana2020tailoring}. The difference originates from the structure of the effective Floquet Hamiltonian: while the triangular lattice hosts only a chiral NNN hopping channel, the driven Kagome lattice simultaneously generates chiral NN and NNN hopping processes. Their independent renormalization under the SR-CMFT reconstructs the effective gauge flux associated with the elementary hopping loops [Eq.~(\ref{eq:flux})], leading to a redistribution of the Berry curvature and a distinct sequence of topological phase transitions. Consequently, the interacting system exhibits significantly fewer topological transitions than the non-interacting Floquet system. The surviving transitions occur only near the zeros of the effective NN hopping, as reflected in the spinon Chern numbers shown in Fig.~\ref{fig-3}(a).

The sublattice-resolved electronic density of states (DOS), computed within the slave-rotor framework (Appendix~\ref{DOSAPP}) and shown in Fig.~\ref{fig:Dos}, provides an independent thermodynamic signature of this correlation-driven reconstruction: the spectral weight at the Fermi level collapses on crossing from the Floquet metal ($U/t=0$) to the correlated insulator ($U/t=4.0$), confirming the opening of a genuine Mott gap.

A further consequence of the correlation-induced band reconstruction is the emergence of topological spinon flat bands at selected driving amplitudes (Fig.~\ref{fig-4}). In particular, the middle spinon band becomes nearly dispersionless [Fig.~\ref{fig-4}(a) and (c)] while retaining a non-trivial topological character, despite the electronic spectrum remaining fully gapped throughout the Mott insulating phase. These nearly flat bands therefore arise as purely collective spinon excitations rather than electronic quasiparticle states. Their coexistence with a robust Mott charge gap establishes the periodically driven Kagome lattice as a promising platform for realizing strongly correlated topological flat-band physics.

\begin{figure}[t]
	\hspace*{\fill}%
	\includegraphics[width=1\linewidth]{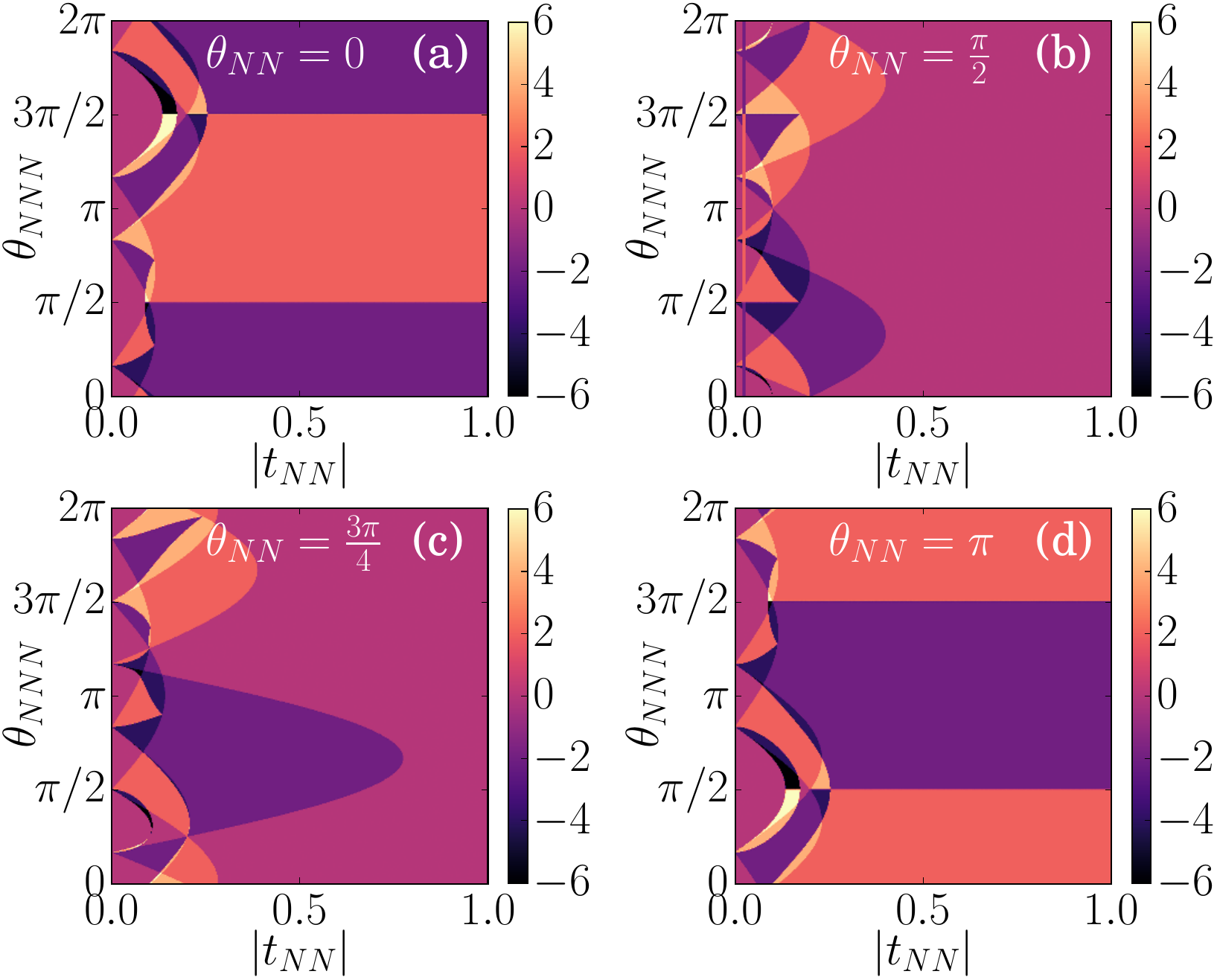}
	\hspace*{\fill}%
	\caption{(Color online) This figure shows the change of the Chern number of the middle band as a function of complex NN (denoted as $t_{NN}$)and NNN (characterised as $t_{NNN}$)hopping. The x-axis is the modulus value of the NN hoping which varies from 0 to 1, and the y-axis is the argument of the NNN hoping (denoted as $\theta_{NNN}$). The modulus of the NNN hopping is fixed at $0.1t$. Four panels from (a) to (d) are for different values of the argument of NN hoping, respectively $\theta_{NN}=0,\frac{\pi}{2},\frac{3\pi}{4}$ and $\pi$.}
	\label{NR}
\end{figure}

Motivated by these results, we propose a minimal effective chiral tight-binding Hamiltonian that captures the essential physics responsible for the rich topological phase diagram and the interaction-driven flat-band engineering reported here,
\begin{equation}
\mathcal{H}_{\mathrm{eff}} = \sum_{\langle i,j\rangle} t_{NN}\, e^{i\theta_{NN}}\, d_i^\dagger d_j + \sum_{\langle\langle i,j\rangle\rangle} t_{NNN}\, e^{i\theta_{NNN}}\, d_i^\dagger d_j + \mathrm{H.c.}
\label{efec}
\end{equation}
in which the NN and NNN hopping amplitudes can be tuned independently in both magnitude and phase. The resulting topological phase diagram (Fig.~\ref{NR}) reveals a rich landscape of Chern insulating phases governed jointly by the relative hopping strengths and their complex phases, with multiple phase boundaries marked by band-gap closings. This effective model unifies a broad class of chiral Kagome Hamiltonians within a single framework and offers a systematic route for navigating this topological phase space via Floquet engineering---providing a blueprint for designing materials or synthetic quantum systems with targeted Chern insulating phases and topological flat bands through light-induced band engineering.

\section{Experimental Feasibility}
\label{sec_IV}

The periodically driven interacting Kagome lattice studied in this work lies within the reach of present-day ultracold atom experiments. Optical Kagome lattices have already been realized using ultracold atoms in tunable optical potentials~\cite{PhysRevLett.108.045305}, where both the lattice geometry and hopping amplitudes can be controlled with high precision. Typical tunneling amplitudes are of the order of
$t\sim100$--$1000\,\mathrm{Hz}$, while onsite interactions can be tuned over a wide range of $U/t$ using magnetic Feshbach resonances~\cite{Jordens2008,Schneider2008,RevModPhys.82.1225}. Such systems provide a natural setting for exploring the interplay between lattice geometry, strong correlations, and nonequilibrium driving.

Periodic driving of optical lattices through lattice shaking or time-dependent synthetic gauge fields has emerged as a powerful route for Floquet engineering of band structures and topological phases~\cite{mikami2016brillouin,PhysRevX.4.031027}. The off-resonant regime considered here requires $\Omega \gg t,U,$
such that the dynamics is well described by the effective Floquet Hamiltonian obtained from the high-frequency expansion. Although isolated periodically driven systems eventually absorb energy and approach an infinite-temperature state, the heating rate is strongly suppressed in the high-frequency regime, leading to a long-lived prethermal window before thermalization sets in~\cite{Abanin2017}. For experimentally relevant parameters satisfying $\Omega/t \gg 1$, this prethermal lifetime can substantially exceed typical experimental observation times, allowing the effective Floquet Hamiltonian to govern the accessible dynamics.

The experimental realization of the Floquet Haldane model using periodically driven ultracold fermions~\cite{Jotzu2014} has already demonstrated that optical lattice shaking can generate complex hopping amplitudes and topological band structures. Similar Floquet protocols could therefore be employed to engineer the drive-induced chiral nearest-neighbour and next-nearest-neighbour hopping processes appearing in the effective Hamiltonian derived in the present work.

Among the most distinctive predictions of our study is the non-monotonic evolution of the Mott transition under periodic driving, leading to repeated metal--insulator transitions as the drive amplitude is varied. In ultracold atomic systems, such transitions may be probed through measurements of compressibility, double occupancy, or momentum-resolved spectroscopy. Furthermore, the continuous tunability of the effective bandwidth and the position of the nearly flat bands through the driving amplitude offers an attractive route for investigating interaction effects in topological flat-band systems away from equilibrium.

Overall, recent advances in Floquet engineering, synthetic gauge fields, and programmable optical lattices suggest that several of the phenomena predicted in this work, including drive-controlled topological transitions, flat-band engineering, and correlation-driven metal-insulator transitions, may be accessible in current or near-future experimental platforms.

\section{Summary and Outlook}
\label{sec_V}

In this work, we investigated the interplay between strong electronic correlations and off-resonant circularly polarized driving in the Kagome Hubbard model using BW perturbation theory and SR-CMFT. Even in the non-interacting limit, the periodic drive not only renormalizes the nearest-neighbour hopping amplitudes but also generates emergent chiral hopping processes that break time-reversal symmetry and induce non-trivial Floquet band topology.

The periodic drive gives rise to multiple band inversions and topological phase transitions while enabling continuous control over the position of the nearly flat band within the spectrum, thereby providing a route toward Floquet engineering of topological flat bands in Kagome systems.

In the presence of Hubbard interactions, we find a non-monotonic evolution of the critical interaction strength for the Mott transition, leading to repeated metal-insulator transitions as the drive amplitude is varied. Within the Mott phase, the spinon sector inherits the drive-induced chirality; however, interaction effects renormalize the different hopping channels unequally. As a consequence, the effective gauge flux experienced by the spinons is modified, leading to spinon band topologies that can differ from those of the underlying electronic Floquet bands.

One notable consequence of this reconstruction is the emergence of topological spinon flat bands in the correlated insulating regime. More generally, our results suggest that in driven correlated systems containing multiple complex hopping channels, electronic correlations may qualitatively modify the topology generated by Floquet engineering rather than simply renormalizing it.

The driven Kagome Hubbard model therefore provides a promising setting for exploring the combined effects of geometric frustration, strong correlations, and nonequilibrium topology. Extensions to multiorbital Kagome materials and related frustrated lattices, as well as investigations of possible interaction-driven phases emerging from partially filled topological flat bands, constitute interesting directions for future work. Recent advances in Floquet engineering, synthetic gauge fields, and programmable quantum simulators further suggest that several of the phenomena discussed here may become experimentally accessible in the near future.

\section{Acknowledgement}
We acknowledge the use of the NOETHER, VIRGO and KALINGA clusters at NISER for numerical computations and fruitful discussion with Anamitra Mukherjee and Ashis Kumar Nandy. RG and SP contributed equally for this work.
\begin{widetext}
\begin{appendix}

\section{Brillouin-Wigner (BW) perturbation} {\label{BW}}
To study the effect of periodic drive in the high frequency limit, we consider the external irradiation of kind $\mathbf{A}=A_0(\cos{(\Omega \tau)},\sin{(\Omega \tau)})$ where $A_0$ and $\Omega$ are the strength and frequency of the drive. The primitive unit vectors are $\mathbf{a}_1=(a,0),\,\mathbf{a}_2=(a/2,\sqrt{3}a/2),\,\mathbf{a}_3=\mathbf{a}_2-\mathbf{a}_1=(-a/2,\sqrt{3}a/2)$ where $a$ is the atomic distance. \\
\bea
 \mathbf{A}.\mathbf{a}_i&=&A_0a(\cos{(\Omega \tau)},\sin{(\Omega \tau)}).(\cos{(\pi l/3)},\sin{(\pi l/3)})\non\\
 &=& A_0a \cos{(\Omega \tau-\pi l/3)}
 \eea

 We begin with a time periodic Hamiltonian $H(\tau + T) = H(\tau)$ given by its Fourier components as
\bea
\mathcal{H}_{p} &=& \int_{0}^{T}\frac{d\tau}{T} H(\tau) e^{ip \Omega \tau}\non\\
&&=\int_{0}^{T}\frac{d\tau}{T} e^{ip \Omega \tau} e^{-iA_0a \sin{(\Omega \tau+\pi/2-\pi l/3)}} t \,[ c_{nm}^{\dagger}a_{nm}+ b_{nm}^{\dagger}c_{nm}+ a_{nm}^{\dagger}b_{nm}\non\\
&&+ c_{n,m-1}^{\dagger}a_{nm}+ b_{n-1,m+1}^{\dagger}c_{nm}+ a_{n+1,m}^{\dagger}b_{nm}]+H.C\non\\
&&=e^{-ip(\pi/2-\pi l/3)}J_p(A_0a) t\,[ c_{nm}^{\dagger}a_{nm}+ b_{nm}^{\dagger}c_{nm}+ a_{nm}^{\dagger}b_{nm}\non\\
&&+ c_{n,m-1}^{\dagger}a_{nm}+ b_{n-1,m+1}^{\dagger}c_{nm}+ a_{n+1,m}^{\dagger}b_{nm}]+H.C\non\\
&&=-t\,[ c_{nm}^{\dagger}a_{nm} e^{-i\pi/6} J_p(A_0)+ b_{nm}^{\dagger}c_{nm}e^{i\pi/6} J_p(-A_0)+ a_{nm}^{\dagger}b_{nm}e^{-i\pi/2} J_p(-A_0)\non\\
&&+c_{n,m-1}^{\dagger}a_{nm}e^{-i\pi/6} J_p(-A_0)+ b_{n-1,m+1}^{\dagger}c_{nm}e^{i\pi/6} J_p(A_0)+ a_{n+1,m}^{\dagger}b_{nm}e^{-i\pi/2} J_p(A_0)]+H.C
\label{H_kag}
\eea

where $T=2\pi/\Omega$ is the period of the drive. We have used the form of the Bessel function (of order p) : 
$J_p(x)=\frac{1}{2\pi}\int_{-\pi}^{\pi} d\tau e^{x\sin{\tau}-p\tau} $\\

Using Brillouin Wigner perturbation theory, we obtain

\bea
H_{BW}&=&\sum_{l=0}^{\infty} H_{BW}^{l}\\
H_{BW}^{(0)}&=&H_0\\
H_{BW}^{(1)}&=&\sum_{p\neq 0}\frac{H_{-p} H_p}{p \Omega}\\
H_{BW}^{(2)}&=&O(\frac{1}{\Omega^2})
\eea
where 
\beq
H_p=\frac{1}{T}\int_0^T d\tau e^{ip \Omega t} H(\tau)
\eeq

$H_0=J_0(A_0) H_{Free}$ : it contains only renormalized free Hamiltonian.\\
\bea
H_p&=&-t( c_{nm}^{\dagger}a_{nm} J_p(A_0)e^{-ip\pi/6}+ c_{n,m-1}^{\dagger}a_{nm} J_{p}(-A_0)e^{-ip\pi/6}\non\\
&&+ b_{nm}^{\dagger}c_{nm}J_p(-A_0)e^{ip\pi/6}+ b_{n-1,m+1}^{\dagger}c_{nm}J_p(A_0)e^{ip\pi/6}\non\\
&&+ a_{nm}^{\dagger}b_{nm}J_p(-A_0)e^{-ip\pi/2}+t_1 a_{n+1,m}^{\dagger}b_{nm}J_p(A_0)e^{-ip\pi/2})+H.C
\eea
and one obtains considering \{$p\neq0$\} : 
\bea
\frac{H_{-p} H_{p}}{p\Omega}&=&-\frac{2i}{p\Omega}\sum_{n,m,\sigma }t^2\,[\{ (c_{nm}^{\dagger}b_{nm}+c_{n,m-1}^{\dagger}b_{n-1,m})+ (b_{nm}^{\dagger}a_{nm}+b_{n-1,m+1}^{\dagger}a_{n,m+1})\non\\
&&+ (a_{nm}^{\dagger}c_{nm}+c_{n+1,m}^{\dagger}c_{n+1,m-1})\}\non\\
&&+\{ (c_{nm}^{\dagger}b_{n-1,m}+c_{n,m-1}^{\dagger}b_{n,m}+c_{n+1,m}^{\dagger}b_{n,m}+c_{n,m}^{\dagger}b_{n,m+1})\non\\
&&+ (b_{nm}^{\dagger}a_{n,m+1}+b_{n-1,m+1}^{\dagger}a_{nm}+b_{nm}^{\dagger}a_{n+1,m-1}+b_{n,m-1}^{\dagger}a_{nm})\non\\
&&+ (a_{nm}^{\dagger}c_{n+1,m-1}+a_{n+1,m}^{\dagger}c_{nm}+a_{n,m}^{\dagger}c_{n-1,m}+a_{n-1,m+1}^{\dagger}c_{nm})\}(-1)^p]\sin{(\frac{\pi p}{3})} J_p(A_0)^2\non\\
\label{H_p}
\eea
\\
Terms inside later \{..\} in Eq.\ref{H_p} can be written like $-i\lambda \sum_{\langle\langle ij\rangle\rangle} \nu_{ij} c_i^\dagger c_j$ \textit{i.e} \{..\} part is similar to NNN SOC. So the final Hamiltonian is 
\beq
H=H_0+\sum_{p=1}^{4} \frac{H_{-p} H_{p}}{p\Omega}
\label{H}
\eeq

\subsection{Hamiltoian in momentum space}

We now write the Hamiltonian in momentum space which is found easily after Fourier transforming Eq. \ref{H}.
\bea
H(k)&=&-2 J_0(A_0) t
\begin{bmatrix}
   0  &  \cos{k_1}  &    \cos{k_2}   \\
       \cos{k_1}    & 0 &   \cos{k_3}  \\
      \cos{k_2}   &   \cos{k_3}  & 0  
\end{bmatrix}\nonumber\\
&&-2i t^2 \sum_{p=1}^{4} \frac{\sin{(\pi p/3)}}{p\Omega} J_p(A_0)^2 
\begin{bmatrix}
   0  &  \cos{k_1}  &   - \cos{k_2}   \\
       - \cos{k_1}    & 0 &   \cos{k_3}  \\
      \cos{k_2}   & - \cos{k_3}  & 0  
\end{bmatrix}\nonumber\\
&&-2i t^2 \sum_{p=1}^{4} (-1)^p \frac{\sin{(\pi p/3)}}{p\Omega} J_p(A_0)^2 
\begin{bmatrix}
   0  & - \cos{(k_2+k_3)}  &   \cos{(k_3-k_1)}   \\
       \cos{(k_2+k_3)}    & 0 &  - \cos{(k_1+k_2)}  \\
     - \cos{(k_3-k_1)}   & \cos{(k_1+k_2)}  & 0  
\end{bmatrix}
\eea
where $k_1=k_x$, $k_2=(k_x+\sqrt{3} k_y)/2$,$k_3=(-k_x+\sqrt{3} k_y)/2$.

\section{Slave-rotor calculation}{\label{SL-formalism}}

Total effective Hamiltonian can be written as 
\begin{equation}\label{A1}
\hat{H}=\sum_{\langle ij\rangle I,J\sigma}(t_{i,j}^Rc_{iI\sigma}^\dag c_{jJ\sigma}+h.c)
+\sum_{\langle ij\rangle I,J\sigma}(t_{i,j}^0c_{iI\sigma}^\dag c_{jJ\sigma}+h.c)
+\sum_{\langle\langle ij\rangle\rangle I,J\sigma}(t_{i,j}^1c_{iI\sigma}^\dag c_{jJ\sigma}+h.c)
+U\sum_{i,I}\hat{n}_{iI\uparrow}\hat{n}_{iI\downarrow}
\end{equation}

Here $t_{i,j}^R$ is the modified bare hopping ,$t_{i,j}^0$ is the emergent nearest neighbor hopping and $t_{i,j}^1$ is the emergent next nearest neighbor hopping due to the periodic drive. $U$ is the onsite Hubbard interaction term.

Now if we substitute the creation and anihilation operator with the following slave-rotor decoupled operators

\begin{eqnarray}\label{a2}
&&c_{jI\sigma}^{\dagger}=f_{jI\sigma}^{\dagger}e^{-i\theta_{jI}},  \
 c_{jI\sigma}=f_{jI\sigma}e^{i\theta_{jI}}
\end{eqnarray}
the Hamiltonian takes the form as 
\begin{multline}
\hat{H}=\sum_{\langle ij\rangle I,J\sigma}(t_{i,j}^Rf_{iI\sigma}^\dag f_{jJ\sigma}e^{-i\theta_{iI}}e^{i\theta_{jJ}}+h.c)
+\sum_{\langle ij\rangle I,J\sigma}(t_{i,j}^0f_{iI\sigma}^\dag f_{jJ\sigma}e^{-i\theta_{iI}}e^{i\theta_{jJ}}+h.c)\nonumber \\
+\sum_{\langle\langle ij\rangle \rangle I,J\sigma}(t_{i,j}^1f_{iI\sigma}^\dag f_{jJ\sigma}e^{-i\theta_{iI}}e^{i\theta_{jJ}}+h.c)
+U\sum_{iI}\hat{n}_{iI\uparrow}\hat{n}_{iI\downarrow}
+\lambda\sum_{iI}(n_{iI}^{\theta}+n_{iI\uparrow}^{f}+n_{iI\downarrow}^{f}-1)
\end{multline}
we assume that the ground $\lvert\Psi\rangle$ state will be the direct product of fermionic ground state $\lvert\Psi^{f}\rangle $ and rotor ground state$\lvert\Psi^{\theta}\rangle$ \textit{i.e}
$\lvert\Psi\rangle=\lvert\Psi^{f}\rangle \lvert\Psi^{\theta}\rangle$. Under this asumption we can decouple the two Hamiltonian as follows $H_{f} \equiv \langle\Psi^{\theta}\lvert H\rvert \Psi^{\theta}\rangle$ and  $H_{\theta} \equiv\langle \Psi^{f}|H|\Psi^{f}\rangle$.
Therefore
\begin{eqnarray}
\hat{H}_f=\sum_{\langle ij\rangle I,J\sigma}(t_{i,j}^Rf_{iI\sigma}^\dag f_{jJ\sigma} \langle e^{-i\theta_{iI}}e^{i\theta_{jJ}}\rangle+h.c)
+\sum_{\langle ij\rangle I,J\sigma}(t_{i,j}^0f_{iI\sigma}^\dag f_{jJ\sigma}\langle e^{-i\theta_{iI}}e^{i\theta_{jJ}}\rangle+h.c) \nonumber\\
+\sum_{\langle\langle ij\rangle \rangle I,J\sigma}(t_{i,j}^1f_{iI\sigma}^\dag f_{jJ\sigma}\langle e^{-i\theta_{iI}}e^{i\theta_{jJ}}\rangle+h.c)
+\mu_f\sum_{iI\sigma}\hat{n}_{iI\sigma}
\end{eqnarray}

\begin{eqnarray}
\hat{H}_\theta=\sum_{\langle ij\rangle I,J\sigma}(t_{i,j}^R\langle f_{iI\sigma}^\dag f_{jJ\sigma}\rangle e^{-i\theta_{iI}}e^{i\theta_{jJ}}+h.c)
+\sum_{\langle ij\rangle I,J\sigma}(t_{i,j}^0\langle f_{iI\sigma}^\dag f_{jJ\sigma}\rangle e^{-i\theta_{iI}}e^{i\theta_{jJ}}+h.c)\nonumber \\
+\sum_{\langle\langle ij\rangle\rangle I,J\sigma}(t_{i,j}^1\langle f_{iI\sigma}^\dag f_{jJ\sigma}\rangle e^{-i\theta_{iI}}e^{i\theta_{jJ}}+h.c)
+\frac{U}{2}\sum_{iI}\hat{n}_{iI}^\theta(\hat{n}_{iI}^\theta-1)
+\mu_\theta\sum_{i}n_{iI}^{\theta}
\end{eqnarray}

We can write the $\hat{H}_f$ in the momentum space as follows

\bea
\hat{H}_f(k)&=&
\begin{bmatrix}
	0  & \tilde{t}\langle e^{-i\theta_{A}}e^{i\theta_{B}}\rangle \cos{k_1}  &   \tilde{t}\langle e^{-i\theta_{A}}e^{i\theta_{C}}\rangle \cos{k_2}   \\
	& 0 &   \tilde{t}\langle e^{-i\theta_{B}}e^{i\theta_{C}}\rangle \cos{k_3}  \\
	&    & 0  
\end{bmatrix}\nonumber\\
&& +
\begin{bmatrix}
	0  &  \tilde{t}^0\langle e^{-i\theta_{A}}e^{i\theta_{B}}\rangle \cos{k_1}  &   \tilde{t}^0\langle e^{-i\theta_{A}}e^{i\theta_{C}}\rangle \cos{k_2}   \\
	& 0 &   \tilde{t}^0\langle e^{-i\theta_{B}}e^{i\theta_{C}}\rangle \cos{k_3}  \\
	&    & 0  
\end{bmatrix}\nonumber\\
&& +
\begin{bmatrix}
	0  &  \tilde{t}^1\langle e^{-i\theta_{A}}e^{i\theta_{B}}\rangle \cos{(k_2+k_3)}  &  \tilde{t}^1\langle e^{-i\theta_{A}}e^{i\theta_{C}}\rangle  \cos{(k_3-k_1)}   \\
	& 0 &   \tilde{t}^1\langle e^{-i\theta_{B}}e^{i\theta_{C}}\rangle \cos{(k_1+k_2)}  \\
	&    & 0  
\end{bmatrix}
\eea
where $k_1=k_x$, $k_2=(k_x+\sqrt{3} k_y)/2$,$k_3=(-k_x+\sqrt{3} k_y)/2$. We have used short form notation :\\
$ \tilde{t}=-2 t J_0(A_0)$,\quad \,$\tilde{t}^0=-2 i t^2 \sum_{p=1}^{4} \frac{\sin{(\pi p/3)}}{p\Omega} J_p(A_0)^2 $,
\quad \,$\tilde{t}^1=2 i t^2 \sum_{p=1}^{4} (-1)^p\frac{\sin{(\pi p/3)}}{p\Omega} J_p(A_0)^2 $, \\

The rotor Hamiltonian is written in real space basis where we have considered 3 sites : a,b,c in an unit cell. To incorporate all the NNN hoppings, we have taken a bigger cluster \textit{i.e}  6 site cluster. Although it seems like that these two Hamiltonians require different size of cluster, there are equivalent points in the rotor cluster.

\subsection{Cluster decomposition of the rotor Hamiltonian}
We consider that the whole system is made of some equivalent cluster. Therefore we can approximate our rotor Hamiltonian as a sum of rotor-cluster Hamiltonian. \textit{i.e} $\hat{H}_\theta=\sum_{I}\hat{H}_I^\theta$. Here $I$ denotes $I^{th}$ cluster.

\begin{multline}
\hat{H}_I^\theta=\sum_{\langle ij\rangle I \sigma}(t_{i,j}^R\langle f_{iI\sigma}^\dag f_{jI\sigma}\rangle e^{-i\theta_{iI}}e^{i\theta_{jI}}+h.c)
+\sum_{\langle ij\rangle I\sigma}(t_{i,j}^0\langle f_{iI\sigma}^\dag f_{jI\sigma}\rangle e^{-i\theta_{iI}}e^{i\theta_{jI}}+h.c)\nonumber \\
+\sum_{\langle\langle ij\rangle\rangle I\sigma}(t_{i,j}^1\langle f_{iI\sigma}^\dag f_{jI\sigma}\rangle e^{-i\theta_{iI}}e^{i\theta_{jI}}+h.c)
+\sum_{\langle ij\rangle  \langle I,J\rangle\sigma}(t_{i,j}^R\langle f_{iI\sigma}^\dag f_{jJ\sigma}\rangle e^{-i\theta_{iI}}\langle e^{i\theta_{jJ}}\rangle+h.c)\nonumber\\
+\sum_{\langle ij\rangle \langle I,J\rangle \sigma}(t_{i,j}^0\langle f_{iI\sigma}^\dag f_{jJ\sigma}\rangle e^{-i\theta_{iI}}\langle e^{i\theta_{jJ}}\rangle+h.c)
+\sum_{\langle\langle ij\rangle \rangle\langle I,J\rangle \sigma}(t_{i,j}^1\langle f_{iI\sigma}^\dag f_{jJ\sigma}\rangle e^{-i\theta_{iI}}\langle e^{i\theta_{jJ}}\rangle+h.c)\nonumber\\
+\frac{U}{2}\sum_{iI}(\hat{n}_{iI}^\theta)^2+\mu_\theta\sum_{iI}n_{iI}^{\theta}\nonumber\\
\end{multline}

We need a 6 site cluster to cover the whole system and incorporate all the hopping.

\begin{figure}[t]
    \centering
    \includegraphics[width=0.7\linewidth]{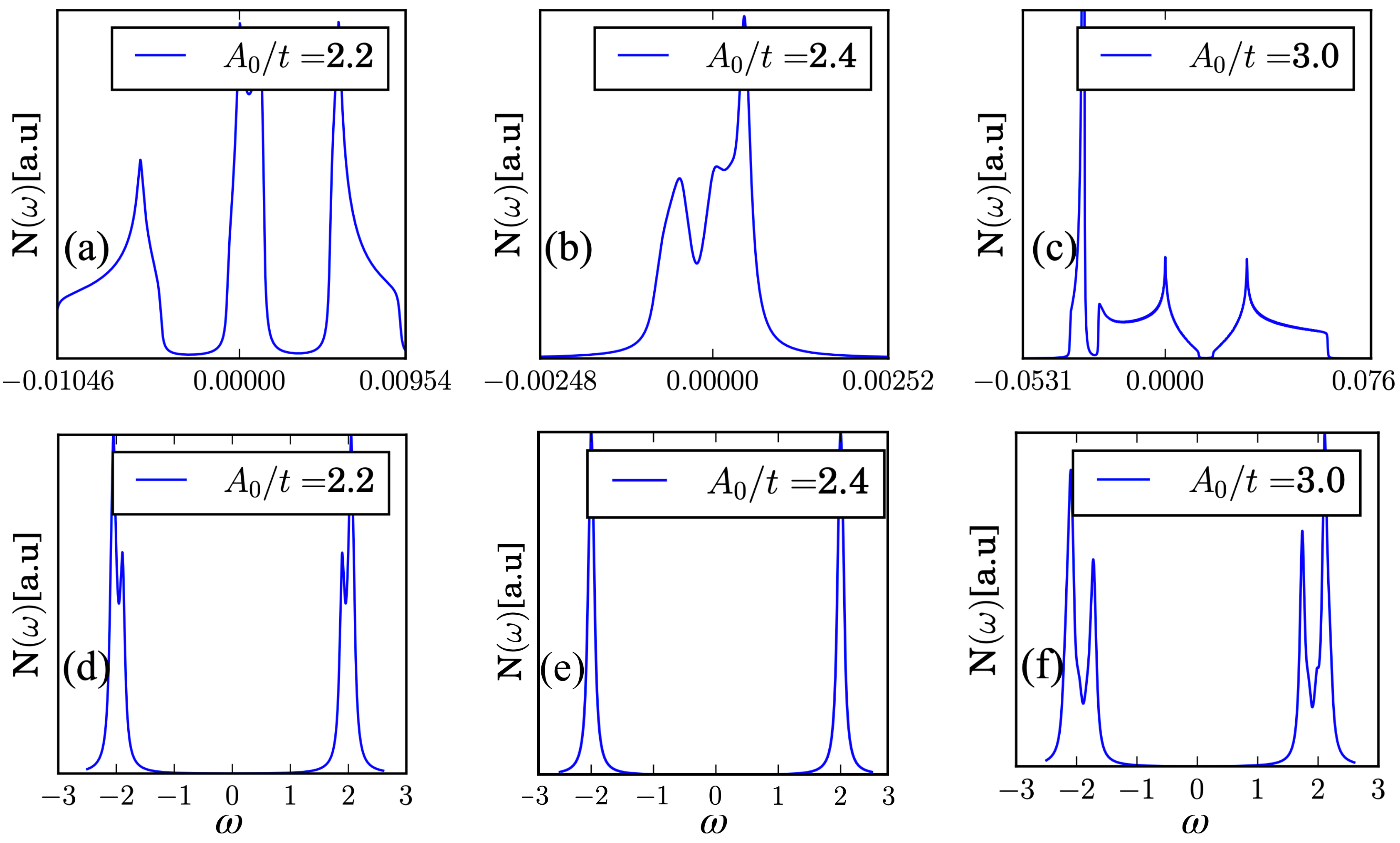}
    \caption{Sublattice-resolved electronic density of states (DOS) calculated within the slave-rotor mean-field framework. Panels (a)--(c) show the non-interacting DOS ($U/t = 0$) for drive amplitudes $A_0 = 2.2$, $2.4$, and $3.0$, respectively, illustrating a robust metallic behavior. Panels (d)--(f) display the corresponding interacting DOS at intermediate interaction strength $U/t = 4.0$ for the same drive amplitudes. The clear suppression of spectral weight at the Fermi energy ($\omega = 0$) dynamically tracks the emergence of a light-tuned topological Mott insulating phase, demonstrating a strongly non-monotonic metal-insulator transition driven by the periodic modulation.}
    \label{fig:Dos}
\end{figure}
\subsection{DOS calculation}{\label{DOSAPP}}
One of the important observable of our calculation is sublattice resolved density of states, defined  as:
\begin{equation}
N_{i}(\omega)=\sum_{\gamma, \sigma} \sum_{I_{i}}\left|\left\langle\chi_{\gamma} \mid I_{i}, \sigma\right\rangle\right|^{2} \delta\left(\omega-\epsilon_{\gamma}\right)
\label{G1}
\end{equation}
Where $i=\{1-6\}$ lattice sites in the $I^{th}$ cluster. Here, ${\left|\chi_{\gamma}\right\rangle}$ and ${\epsilon_{\gamma}}$ are eigenvectors and eigenvalues of $H$. In the slave-rotor formalism we
have decomposed the electron oparator into a rotor and a spinon at every site. There fore the PDOS will be convolution of the rotor DOS and spinon DOS.\\
First we reconstruct the (electron) single-particle Green’s function. The imaginary part of the Green's function is the spectral function and which is the  PDOS. The local retarded Matsubara Green’s function is defined as:
\begin{equation}
\begin{aligned}
G_{iI \sigma}\left(i \omega_{n}\right)=&-\int_{0}^{\beta} d \tau e^{i \omega_{n} \tau}\left\langle\Psi\left|c_{iI, \sigma}(\tau) c_{iI, \sigma}^{\dagger}(0)\right| \Psi\right\rangle \\
=&-\int_{0}^{\beta} d \tau e^{i \omega_{n} \tau}\left\langle\Psi^{f}\left|f_{iI \sigma}(\tau) f_{iI \sigma}^{\dagger}(0)\right| \Psi^{f}\right\rangle 
 \times\left\langle\Psi^{\theta}\left|e^{-i \theta_{iI}(\tau)} e^{i \theta_{iI}(0)}\right| \Psi^{\theta}\right\rangle
\end{aligned}
\label{G2}
\end{equation}
The
spinon correlator in Eq. \ref{G2} is calculated as:
\begin{equation}
\begin{array}{l}
\frac{1}{2} \sum_{\sigma}\left\langle f_{iI \sigma}(\tau) f_{iI \sigma}^{\dagger}(0)\right\rangle
\quad=\frac{1}{2} \sum_{\gamma \sigma}\left|\left\langle\chi_{\gamma}^{f} \mid iI, \sigma\right\rangle\right|^{2}\left[1-n_{f}\left(\epsilon_{\gamma}^{f}-\mu_{f}\right)\right] e^{-\tau\left(\epsilon_{\gamma}^{f}-\mu_{f}\right)}
\end{array}
\label{G3}
\end{equation}
Where $\left|\chi_{\gamma}^{f}\right\rangle$ and $\epsilon^f_{\gamma}$ are the spinon eigenvectors and eigenvalues, respectively. The rotor correlator in Eq. \ref{G2} is
\begin{equation}
\begin{array}{l}
\left\langle e^{-i \theta_{iI, \sigma}(\tau)} e^{i \theta_{iI, \sigma}(0)}\right\rangle
=\frac{1}{Z_{\theta}} \sum_{m, n} e^{-\beta \epsilon_{m}}\left\langle m\left|e^{-i \theta_{iI, \sigma}}\right| n\right\rangle\left\langle n\left|e^{i \theta_{iI, \sigma}}\right| m\right\rangle e^{\tau\left(\epsilon_{m}-\epsilon_{n}\right)},
\end{array}
\label{G4}
\end{equation}
Where $\epsilon_m$ and $|m\rangle$ are the eigenvalues and corresponding eigenvectors of the rotor Hamiltonian. Here, $Z_{\theta}$  is the rotor partition function defined as:
\begin{equation}
Z_{\theta}=\sum_{m} e^{-\beta \epsilon_{m}}
\label{G5}
\end{equation}
In the Eq. \ref{G2} integrating over imaginary time $\tau$ and then taking the analytically continue we obtained $G_{iI \sigma}(\omega)$. The PDOS is obtained from its imaginary part of $G_{iI \sigma}(\omega)$.

To systematically monitor the dynamic reconstruction of the electronic structure and map out the phase boundary of the light-tuned topological Mott phase, we compute the sublattice-resolved electronic density of states (DOS), as illustrated in Fig.~\ref{fig:Dos}. In the non-interacting limit ($U/t=0$), panels (a)--(c) exhibit a robust, finite spectral weight across the Fermi energy ($\omega = 0$), characterizing the ideal Floquet metallic state where the flat band is systematically driven across the spectrum via the drive amplitude $A_0$. Upon turning on intermediate electronic correlations ($U/t=4.0$), the system undergoes a striking structural reorganization: the spectral weight at $\omega=0$ becomes heavily suppressed, culminating in the opening of a clean correlation-driven charge gap [Figs.~\ref{fig:Dos}(d)--(f)]. Crucially, this evolution directly visualizes the non-monotonic nature of the Mott transition under external driving.

\end{appendix}
\end{widetext}

\bibliography{bibfile} 

\end{document}